\documentclass[final]{cvpr}

\usepackage{times}
\usepackage{epsfig}
\usepackage{graphicx}
\usepackage{amsmath}
\usepackage{amssymb}

\usepackage{color}
\usepackage{multirow}
\usepackage{bbding}
\usepackage{subfigure}
\usepackage[pagebackref=true,breaklinks=true,colorlinks,bookmarks=false]{hyperref}

\def\cvprPaperID{47} 
\def\confYear{CVPRW 2021}

\begin{document}

\title{Anchor-based Plain Net for Mobile Image Super-Resolution}

\author{Zongcai Du \quad Jie Liu \quad Jie Tang\thanks{Corresponding author} \quad Gangshan Wu\\
State Key Laboratory for Novel Software Technology, Nanjing University, China\\
{\tt\small {\{151220022,jieliu\}@smail.nju.edu.cn}, \{tangjie,gswu\}@nju.edu.cn}
}

\maketitle

\begin{abstract}
   Along with the rapid development of real-world applications, higher requirements on the accuracy and efficiency of image super-resolution (SR) are brought forward. Though existing methods have achieved remarkable success, the majority of them demand plenty of computational resources and large amount of RAM, and thus they can not be well applied to mobile device. In this paper, we aim at designing efficient architecture for 8-bit quantization and deploy it on mobile device. First, we conduct an experiment about meta-node latency by decomposing lightweight SR architectures, which determines the portable operations we can utilize. Then, we dig deeper into what kind of architecture is beneficial to 8-bit quantization and propose anchor-based plain net (ABPN). Finally, we adopt quantization-aware training strategy to further boost the performance. Our model can outperform 8-bit quantized FSRCNN by nearly 2dB in terms of PSNR, while satisfying realistic needs at the same time. Code is avaliable at \url{https://github.com/NJU-Jet/SR_Mobile_Quantization}.
\end{abstract}

\section{Introduction}
Single image super-resolution (SISR) is a classical and long-standing problem in low-level computer vision. The goal is to reconstruct a high-resolution (HR) image according to its degraded low-resolution (LR) counterpart. It has been applied widespreadly in multiple diverse fields, such as HDTV~\cite{HDTV}, magnetic resonance imaging~\cite{mri1, mri2}, satellite sensor image reconstruction~\cite{Satellite1, Satellite2}, and underwater applications\cite{Underwater1, Underwater2}. The difficulty of SISR is that numerous HR images can map to an identical LR image even through the same degradation model. To find a relatively satisfactory result in the infinite solution space, plenty of traditional SISR algorithms have been proposed in the literature, including but not limited to, interpolation-based~\cite{interpolation1, interpolation2}, image statistics-based~\cite{statistics1, statistics2}, patch-based~\cite{patch1, patch2, patch3} and example-based ~\cite{example1, example2, example3} methods.

Since the dawn of deep learning, CNN-based methods have made further progress in SISR. SRCNN~\cite{SRCNN} innovatively employed a three-layer CNN to directly learn the mapping function and led to significant improvements compared with conventional methods.  After that, more and more creative ideas are introduced, such as residual learning~\cite{VDSR, SRResNet, EDSR}, feature fusion~\cite{RDN, SRFBN, MemNet}, well-designed loss function\cite{SRResNet, EDSR, EnhanceNet} and attention mechanism~\cite{RCAN, SAN, IMDN, LatticeNet}, advancing the performance of SISR.

In recent years, the communities have noticed the deployment issue on mobile device. Most superior models are designed for desktop purposes so they can not be directly applied to mobile environment. In order to strive towards the ultimate goal of applying SISR technology to real-world applications where computational resources are limited, image restoration on smart-phone contests ~\cite{contest1, contest2, contest3} have been held to shed a light upon this problem. Meanwhile, most mobile devices are embedded with deep learning accelerators, and some benchmark suites~\cite{benchmark1, benchmark2} are developed to measure their performances. For creating mobile-friendly models, there are two basic ideas. One is toward network optimization which can be mainly categorized into quantization~\cite{quantization1, quantization2}, pruning~\cite{pruning1, pruning2} and knowledge distillation~\cite{distillation1, distillation2}. The other is toward lightweight architecture design\cite{IDN, IMDN, CARN, LatticeNet}. Our focus is how to create a general SISR network architecture which is beneficial to 8-bit quantization.

In spite of achieving prominent improvements, yet there are drawbacks in two aspects. First, the obtained models are usually evaluated on desktop CPUs and GPUs, making it nearly impossible to estimate the actual inference time and memory consumption on real mobile hardware. Second, even recent state-of-the-art (SOTA) lightweight models include dozens of convolution nodes~\cite{EDSR, CARN, IDN} and time-consuming nodes such as attention~\cite{IMDN, LatticeNet}, making them impracticable for realistic use-cases (\eg restore twenty-four 1080P video frames per second). We resolve the drawbacks by researching meta-node latency on mobile hardware and digging deeper into what kind of architecture can really make sense to INT8 quantization. In summary, our main contributions are as follows:
\begin{itemize}
	\item We investigate meta-node latency on mobile hardware according to SOTA lightweight SISR architectures, which yields portable operations.
	\item We propose anchor-based residual learning strategy, which is much faster than nearest neighbor resize on mobile device, and can largely improve the INT8 quantized model performance by nearly 2dB without any parameter cost.
	\item We propose anchor-based plain net (ABPN) for mobile SISR, which is able to restore twenty-seven 1080P images (x3 scale) per second, maintaining good perceptual quality as well.
\end{itemize}

\section{Related Work}
\subsection{Overview of image super-resolution}
Recently, deep learning based methods have achieved dramatic improvements in various kinds of tasks including SISR. Dong \etal~\cite{SRCNN} innovatively introduced a deep learning model called SRCNN to reconstruct HR image in an end-to-end manner. Although SRCNN outperforms hand-crafted models by a large margin, it entails high computational loads due to learning in the HR space. To solve the problem, shi \etal proposed ESPCN~\cite{ESPCN} to replace the bicubic filter with a more efficient sub-pixel convolution. In the same period, Kim \etal~\cite{VDSR} deepened the network to twenty layers, indicating that the depth is crucially important for SISR task. Subsequently, Ledig \etal~\cite{SRResNet} introduced residual block (RB)~\cite{ResNet} to maximize the power of residual learning. Furthermore, based on SRResNet~\cite{SRResNet}, Lim \etal~\cite{EDSR} presented an enhanced deep super-resolution network (EDSR), which made a breakthrough by removing unnecessary modules in RB and had a far-reaching impact on the succeeding studies~\cite{RDN, RCAN, CARN, RNAN, RFA, LatticeNet}. For example, RDN~\cite{RDN} proposed residual dense block to make full use of all the hierarchical features via dense connected convolution layers. RCAN~\cite{RCAN} integrated channel attention mechanism into RB and adopted residual-in-residual structure to form a very deep network.

\subsection{Lightweight image super-resolution}
Due to the growing realistic demand, many works have devoted to making models lighter and faster. They can be divided into explicit~\cite{DRCN, DRRN, SRFBN, CARN, IDN} and implicit schemes~\cite{LapSRN, CARN, MemNet, IMDN, LatticeNet}. The former adopts simple operations to explicitly reduce the model complexity, such as directly cutting down the width and depth~\cite{SRFBN}, recurrent structure~\cite{DRCN, DRRN} and group convolution~\cite{CARN, IDN}. These naive strategies bring about either accuracy loss or more extra overheads (\eg, FLOPs). The latter implicit scheme concentrates on sufficiently utilizing intermediate features as well as enhancing the discriminative capability of the network, thus leading to less computational cost and better results on the whole. For instance, LapSRN~\cite{LapSRN} exploited features at each pyramid level to restore the sub-band residuals of different high-resolution images. MemNet~\cite{MemNet} introduced gating mechanism to bridge the long-term with short-term information. CARN~\cite{CARN} implemented cascading mechanism to incorporate features at both local and global level. IMDN~\cite{IMDN} retained partial information as refined features and fused the distilled features  by contrast-aware channel attention (CCA) mechanism. LatticeNet~\cite{LatticeNet} created a butterfly structure and also applied CCA to dynamically combine two RBs. The capacity of lightweight models is limited, so recent architecture designs pay attention to making full use of information of different levels as has been stated above. However, situation on mobile hardware is totally different from desktop CPUs and GPUs. For example, hierarchical feature fusion~\cite{CARN, IDN, IMDN, LatticeNet} would cause slow access to RAM due to limited cache memory on mobile device. Another popular strategy, attention mechanism~\cite{RCAN, SENet, IMDN, LatticeNet}, would also lead to unbearable time overhead because of calculating global statistics and element-wise multiplication.

\subsection{Network Quantization}
Quantization is a process of distributing continuous real-valued infinite numbers to a smaller set of discrete finite values, for minimizing the number of bits required and also maximizing the accuracy of the attendant computations. The widely used full-integer quantization technology can be three times faster than original float-point network, and holds the potential to reduce memory footprint by a factor of 4x. Post-training quantization and quantization-aware training are two famous techniques supported by TensorFlow~\cite{tensorflow}. The former estimates value ranges of network parameter and activation by traversing provided representative data after training, while the latter finish this process during training by inserting fake quantization nodes. Both of them have demonstrated promising results on image perception tasks~\cite{task1, task2, task3}, but applying them to SR task is much harder and will incur significant accuracy drop. The reason is that current architectures remove batch normalization (BN) layers because they result in blurred reconstructed HR images with artifacts~\cite{EDSR}, but the removal also leads to high dynamic quantization range at the same time. There are limited works~\cite{Q1, Q2} targeting on solving this problem. ~\cite{Q1} binarizes the convolution filters only in residual blocks, and adopts a learnable weight for each binary filter, which can not be applied to full-integer device. ~\cite{Q2} proposes PArameterized Max Scale to explore the upper bound of the quantization range adaptively, which increase training complexity due to manually selecting hyper-parameter and structured knowledge transfer. Simple and useful technology still remains to be explored.

\section{Proposed Method}
\begin{figure*}[t]
	\centering
	\includegraphics[width=1.0\linewidth]{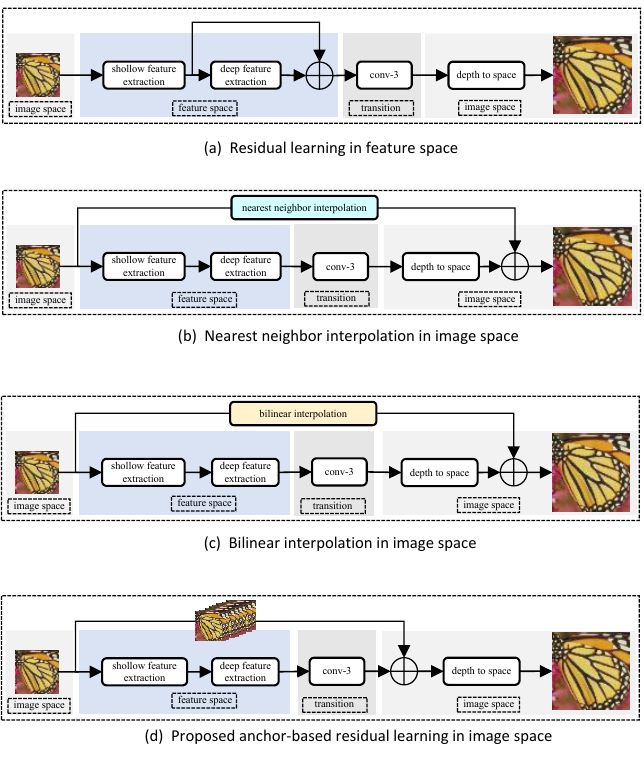} 
	\caption{Illustration of residual learning in image space and feature space.}
	\label{fig:RL}
\end{figure*}

\begin{figure*}[t]
	\centering
	\includegraphics[width=1.0\linewidth]{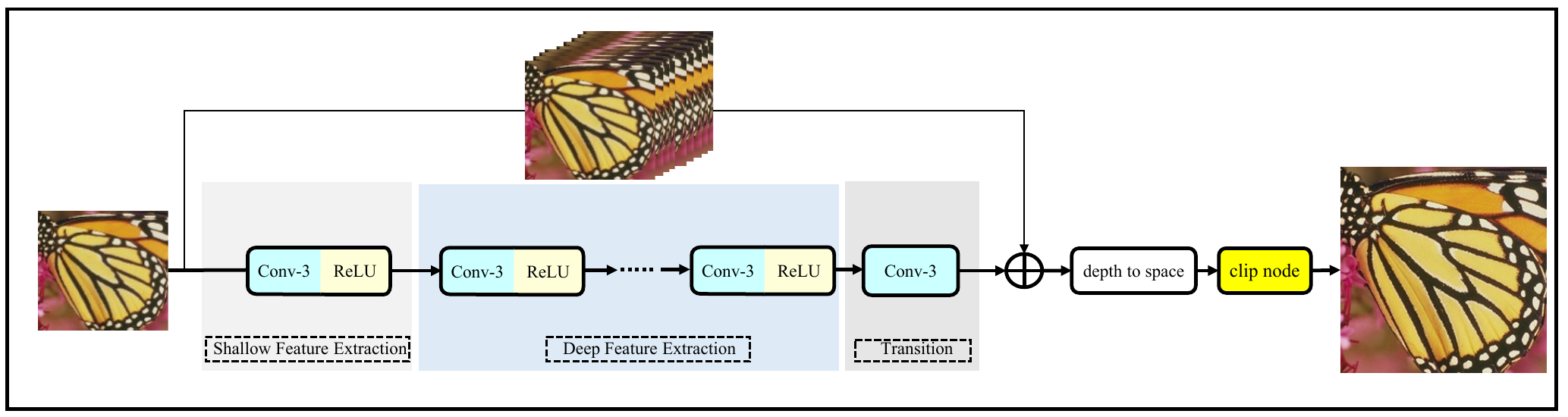} 
	\caption{The whole network architecture.}
	\label{fig:architecture}
\end{figure*}
In this section, we start from investigating meta-node latency. Then, we introduce the insight behind our anchor-based residual learning. Finally, we build our final ABPN and describe the design principle of each component.

\subsection{Meta-node Latency}
\label{Meta-node Latency}
We bear in mind that our goal is to create a real-time model qualified for realistic use-cases (\eg super-resolve video frames). The first thing is figuring out the set of portable meta-node and time-consuming meta-node. We create our initial meta-node set by decomposing recent lightweight SR architectures~\cite{EDSR, CARN, IDN, IMDN, LatticeNet} and then test these meta-nodes on Synaptics Dolphin Platform with a dedicated NPU. They can be divided into four categories: tensor operator nodes, convolution nodes, activation nodes and resize nodes. From Table.~\ref{tab:meta-nodes}, we have four observations. First of all, recent technologies used in SOTA lightweight architectures seem to be impracticable to be deployed on mobile device. EDSR~\cite{EDSR} adopts a mass of RBs, and each RB will introduce an element-wise addition which is even slower than highly-optimized convolution layer. CARN incorporates global and local features, and each incorporation includes one concatenation of large amount of channels and one $1\times1$ convolution, bringing about only 0.09dB improvement according to their article. IDN~\cite{IDN} and IMDN~\cite{IMDN} are also in dire straits on mobile device, for rapid feature split and concat. It's more serious for LatticeNet~\cite{LatticeNet} which adopts sixteen CA blocks, and each CA block contains one element-wise addition and multiplication, two pooling layers, and four $1\times1$ convolution layers. Total sixteen CA blocks can only contribute to 0.15dB improvement while leading to heavy computational burden. Another common problem is that they all need to retain features of previous layers, and utilize $1\times1$ convolution layer to control how much of the previous states should be reserved, and determine how much of the current state should be stored. This long-term dependency causes frequent slow contaction with RAM since there is only limited cache memory in mobile device. Thus, we would not take feature fusion, feature distillation, group convolution and attention mechanism into consideration. Second, although the number of parameters and floating point operations of $3\times3$ convolution layer is nine times as large as that of $1\times1$ convolution layer, the time consumption does not differ much due to parallel calculation. So we prefer to utilize $3\times3$ convolution layer to produce larger receptive fields, which is critically important for  micro-architecture. Third, as for activation function, we choose ReLU because it is much faster than Leaky ReLU and we find that the performance gain of Leaky ReLU is quite small (within 0.03dB). Last, resize nodes are too slow because of coordinate mapping between interpolated HR image and input LR image.

\begin{table}[!htbp]
	\centering
	\caption{Meta-nodes inference time (ms) on Synaptics Dolphin Platform. Resize nodes are applied to network input, while other nodes are applied to 1080P network output.}
	\smallskip
	\label{tab:meta-nodes}
	\resizebox{1.0\linewidth}{!}{
		\begin{tabular}{|c|c|c|c|c|c|c|}
			\hline
			\multicolumn{3}{|c}{Main type} & \multicolumn{3}{|c|}{ Meta-node} & Time\\
			\hline
			\hline
			\multicolumn{3}{|c|}{\multirow{6}{*}{Tensor operator nodes}} & \multicolumn{3}{c|}{Channel split} & 9.8 \\
			\cline{4-7}
			\multicolumn{3}{|c|}{} & \multicolumn{3}{c|}{Channel concat} & 10.4 \\
			\cline{4-7}
			\multicolumn{3}{|c|}{} & \multicolumn{3}{c|}{Add two tensors} & 5.2 \\	
			\cline{4-7}
			\multicolumn{3}{|c|}{} & \multicolumn{3}{c|}{Multiply two tensors} & 9.6 \\
			\cline{4-7}
			\multicolumn{3}{|c|}{} & \multicolumn{3}{c|}{Global max pooling} & 20.0 \\
			\cline{4-7}
			\multicolumn{3}{|c|}{} & \multicolumn{3}{c|}{Global average plooling} & 13.1 \\				
			\hline
			\hline
			\multicolumn{3}{|c|}{\multirow{2}{*}{Convolution nodes}} & \multicolumn{3}{c|}{$3\times3$ Convolution} & 4.3 \\
			\cline{4-7}
			\multicolumn{3}{|c|}{} & \multicolumn{3}{c|}{$1\times1$ Convolution} & 2.9 \\
			\hline	
			\hline	
			\multicolumn{3}{|c|}{\multirow{2}{*}{Activation nodes}} & \multicolumn{3}{c|}{ReLU} & 1.3 \\
			\cline{4-7}
			\multicolumn{3}{|c|}{} & \multicolumn{3}{c|}{Leaky ReLU} & 3.6 \\
			\hline
			\hline
			\multicolumn{3}{|c|}{\multirow{2}{*}{Resize nodes}} & \multicolumn{3}{c|}{Nearest neighbor} & 57.6 \\
			\cline{4-7}
			\multicolumn{3}{|c|}{} & \multicolumn{3}{c|}{Bilinear} & 75.4 \\
			\hline							
		\end{tabular}}	
\end{table}

\subsection{Anchor-based Residual Learning}
\label{ABRL}
As has been discussed in Sec.~\ref{Meta-node Latency}, the available meta-nodes are really limited. To get a good solution, we need to dig deeper into the relationship between architecture design and INT8 quantization. As we know, the difficulty lies in high dynamic range of image-to-image mapping, so the direct idea is to produce lower standard deviation weights and activations. There are two simple ways to achieve this goal. One is adding BN layer, and the other is residual learning. On the one hand, BN is always integrated into RB, so the introduction will not only induce extra time and memory overhead, but also significantly decrease the performance by about 0.2dB. On the other hand, neighboring pixels always have nearly the same values so it seems nature to learn residual, which is close to zero. Residual learning can be divided into image-space residual learning (ISRL) and feature-space residual learning (FSRL). ISRL is adopted in early works~\cite{VDSR, DRRN} to map a LR image to a blurred HR image, while FSRL is widely adopted in recent SOTA models~\cite{SRResNet, IDN, IMDN} which sightly outperforms ISRL in floating-point space. However, we argue that ISRL is better for INT8 quantization because it forces the whole network to learn small residual, and this intuition will be experimentally verified in Sec.~\ref{Experiment}. From Table~\ref{tab:meta-nodes}, we can see that both image space interpolations suffer from unbearable time cost and even just one single node can not satisfy realistic demand. We recognize it is the floating calculation in coordinate mapping that restricts the deployment of ISRL. To tackle this problem, we propose anchor-based residual learning (ABRL). Different from nearest neighbor interpolation which needs floating calculation in coordinate mapping, ABRL directly repeats every pixel nine times in LR space to generate anchors for every pixel in HR space. Thanks to unique pixel shuffle layer, our ABRL can be easily realized by one channel-concat and one addition meta-node. Fig.~\ref{fig:RL} shows the four different kinds of residual learning strategy. As for time overhead, residual learning in feature space contains only one element-wise addition, which costs 5.2ms. Our proposed ABRL contains one channel-concat and one element-wise addition, which total costs 15.6ms, nearly a quarter of the time cost of nearest neighbor meta-node. It should be noted that the cost of nine channel-concats in LR image space is almost equal to that of one channel-concat in HR image space. Our ABRL has two main advantages: the first is that it can largely improve the performance of INT8 quantized model compared with that of residual learning in feature space (up to 0.6dB); the second is that multi-branch architecture can be inferred parallelly so the actual cost of ABRL is the same as feature-space residual learning, and the main cost of our ABRL and FSRL is brought by slow access to RAM. It is also worth mentioning that our ABRL is a special case of nearest neighbor interpolation when the scale factor is integer.

\subsection{Network Architecture}
The whole architecture is depicted in Fig.~\ref{fig:architecture}. Our architecture mainly consists of four parts: shallow feature extraction part which transfers the image to feature space, deep feature extraction which extracts high-level information and restore details (edges, textures) step by step, reconstruction part which maps features to HR image space, and post  processing part which re-arrange pixels and restricts the values within normal image range.
Let's denote $I_{LR}$ and $I_{SR}$ as the input and output of our ABPN. We obtain shallow feature $F_0$ by:
\begin{equation}
    F_{0} = H_{SFE}(I_{LR})
\end{equation}
where $H_{SFE}\left( \cdot \right)$ denotes the mapping function in shallow feature extraction. We use one $3\times3$ convolution layer followed by ReLU to form this part. After that, we use Conv-ReLU pairs, which is the fastest combination in Table.~\ref{tab:meta-nodes}, to gradually refine details. The $i$-th deep features $F_{i}$ is obtained through:
\begin{equation}
    F_{i} = H_{DFE_{i}}(F_{i-1}), i=1,\dots,5,
\end{equation}
where $H_{DFE_{i}}\left( \cdot \right)$ represents $i$-th Conv-ReLU pair in deep feature extraction part.To fully take advantage of parallel inferring, we set the number of Conv-ReLU pairs to 5 to match the overhead in the upper branch, which means when Conv-ReLU pairs is less 5, the mobile inference time remains the same. Then, one convolution layer is adopted to transfer features to HR image space:
\begin{equation}
    F_{t} = H_{T}(F_{5})
\end{equation}
where $H_{T}\left( \cdot \right)$ is the mapping function in transition layer and $F_{t}$ is the obtained residual image features.Our ABRL is applied subsequently to get the super-resolved image of which the spatial pixels are put in the channel axis:
\begin{equation}
    F_{SR} = F_{t} + I_{LR}
\end{equation}
Finally, pixel shuffle layer is used to re-arrange $F_{SR}$ and a clip node is used to restrict values to get $I_{SR}$:
\begin{equation}
    I_{SR} = H_{PP}(F_{SR})
\end{equation}
where $H_{PP}\left( \cdot \right)$ is the mapping function in post processing part. Clip node, at the tail of the network,clips values less than zero or larger than 255. The absence of this node will cause the shift of output distribution, and when applying full-integer quantization the converter would think that there are negative values of real images.
\subsection{Loss Function}
 There are a lot of loss functions which have been adopted in previous works~\cite{EDSR, IDN, IMDN}. To make sure the improvement is mainly from our design, we simply use $L_{1}$ loss function to optimize our network which can be formulated as:
\begin{equation}
	L(\Theta) = \frac{1}{N}\sum_{i=1}^{N}\left \| f_{ABPN}(I_{LR}^{i}) - I_{HR}^{i} \right \|_{1}
\end{equation}
where $\Theta$ denotes the parameters of our network and $N$ is the total number of training samples. $I_{LR}^{i}$ and $I_{HR}^{i}$ denote the $i$-th LR patch and the corresponding ground truth. $f_{ABPN}\left( \cdot \right)$ represents the operations of the proposed ABPN.
\section{Experiments}
\label{Experiment}
\subsection{Settings}
\textbf{Implementation details.} In each training batch, 16 cropped $64 \times 64$ LR RGB patches augmented by random flipping and rotation are sent to the network. The learning rate is initialized as $1\times 10^{-3}$ and decreases half per 200 epochs for 1000 epochs. The number of kernels in the residual learning branch is set to 28. Parameters of our model is initialized using the method proposed by He \etal~\cite{kaiming} and optimized by ADAM optimizer~\cite{ADAM} with $\beta_1=0.9$, $\beta_2=0.999$, and $\epsilon=10^{-8}$. We follow the Mobile AI image super-resolution challenge~\cite{MAI2021} to measure SR results in the RGB space, and adopt DIV2K~\cite{DIV2K} as training and validation sets.

\subsection{Residual Learning}
In this section, we verify the efficiency of residual learning and the superiority of our ABRL. We first remove the upper branch in Fig.~\ref{fig:architecture} to build our baseline model. Then, we separately add four residual learning strategies to the baseline model. The results are reported in Table.~\ref{tab:ablation}, from which we have the following observations. For FP32 model, FSRL model can achieve the best performance (+0.03dB), while other methods achieves nearly the same performance. For INT8 quantization model, architecture without residual learning suffer from severly accuracy drop (-1.93dB), while the architectures with feature-space residual learning drops 0.78dB and architectures with image-space residual learning drop only 0.13dB. Thus, we can conclude that residual learning can largely allevate high dynamic range problem in INT8 quantization, and image-space residual learning is much better than feature-space residual learning.
\begin{table}[h]
	\centering
	\caption{Investigation of image space and feature space residual learning. The inference time is measured by Synaptics smart TV platform. FSRL denotes feature-space residual learning in Fig.~\ref{fig:RL}, ABRL denotes our proposed anchor-based residual learning.}
	\smallskip
	\label{tab:ablation}
	\resizebox{1.0\linewidth}{!}{
		\begin{tabular}{|c|c|c|c||c|}
			\hline
			Model & Params & FP32 & INT8 & Inference time \\
			\hline
			\hline
			Baseline & 42.54K & 30.21 & 28.28 & 26ms \\
			\hline
			\hline
			Baseline+nearest & 42.54K & 30.22 & 30.09 & 57.6ms \\
			\hline	
			Baseline+bilinear & 42.54K & 30.24 & 30.11 & 74.9ms \\
			\hline	
			Baseline+FSRL & 42.54K & 30.27 & 29.49 & 35.9ms \\
			\hline	
			Baseline+ABRL & 42.54K & 30.22 & 30.09 & 36.8ms \\
			\hline					
	\end{tabular}}
\end{table}

\subsection{Quantize-Aware Training}
Quantization-aware training (QAT) is a popular technology to boost the performance without any inference stage cost. We set initial learning rate to $1\times 10^{-4}$ and decreases half per 50 epochs for 200 epochs. We can further improve the performance by 0.06dB. By now, the INT8 quantized network only lose 0.07dB on mobile image super-resolution compared with its floating point version.

\subsection{Test on Snapdragon 820}
We report the inference time of our model on a real mobile device with Snapdragon 820. We use AI Benchmark\footnote{https://ai-benchmark.com/download} to get the CPU, GPU, NNAPI running time. The results are shown in Table.~\ref{tab:820}.

\begin{table}[h]
	\centering
	\caption{Inference time (ms) of ABPN on Snapdragon 820.}
	\smallskip
	\label{tab:820}
	\resizebox{0.8\linewidth}{!}{
		\begin{tabular}{|c|c|c|c||c|}
		\hline
		Model & CPU & GPU & NNAPI & PSNR \\
		\hline
		FSRCNN & 149.2 & 44.3 & 21.7 & 28.1\\
		\hline
		ABPN & 235.0 & 69.8 & 39.2 & 30.15\\
		\hline
	\end{tabular}}
\end{table}

\subsection{MAI2021 SISR Challenge}
This work is initially proposed for the purpose of participating
in the MAI2021 Single Image Super-Resolution Challenge~\cite{MAI2021}.
We report the preliminary results and our result in Table.~\ref{tab:MAI2021}. Our first submission is the same model without the clip node at the tail of the network, and the results is really bad (less than 20dB). We solve this issue after the deadline and send the corrected model to the organizers. Beneficial from our anchor-based residual learning, our models can outperform other models by a large margin especially for PSNR metric. Also, we can achieve the fastest inference speed.
\begin{table}[h]
	\centering
	\caption{Comparison of our results and official preliminary results on MAI2021 Single Image Super-Resolution track.}
	\smallskip
	\label{tab:MAI2021}
	\resizebox{1.0\linewidth}{!}{
		\begin{tabular}{|c|c|c|c|c|}
			\hline
			& PSNR & SSIM & NPU Runtime & Score \\
			\hline
			ABPN (\textbf{Ours}) & \color{red}{29.87} & \color{red}{0.8686} & \color{red}{36.89} & \color{red}{92.72} \\
			\hline
			deepernewbie & 29.58 & 0.86 & 44.85 & 51.02 \\
			\hline
			JeremieG & 29.41 & 0.8537 & 38.32 & 47.18 \\
			\hline
			richlaji & 29.52 & 0.8607 & 62.25 & 33.82 \\
			\hline
			xindongzhang & 28.82 & 0.8428 & 76.61 & 10.41 \\
			\hline	
	\end{tabular}}
\end{table}

\subsection{Visual Comparision}
\begin{figure*}[!htbp]
		\begin{center}
			\begin{tabular}{@{}c@{}}
                \includegraphics[width=0.9\linewidth]{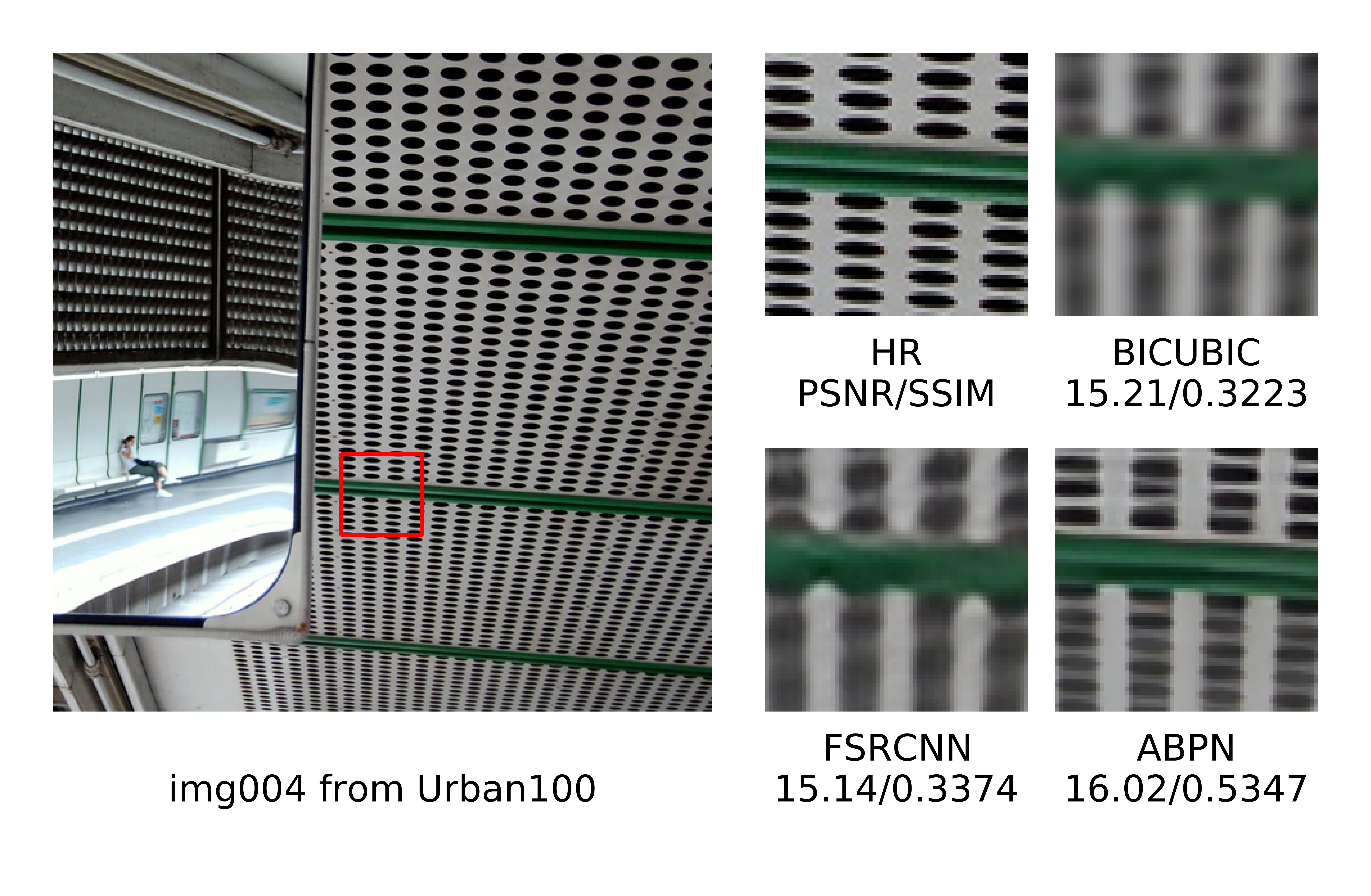} \\
                \includegraphics[width=0.9\linewidth]{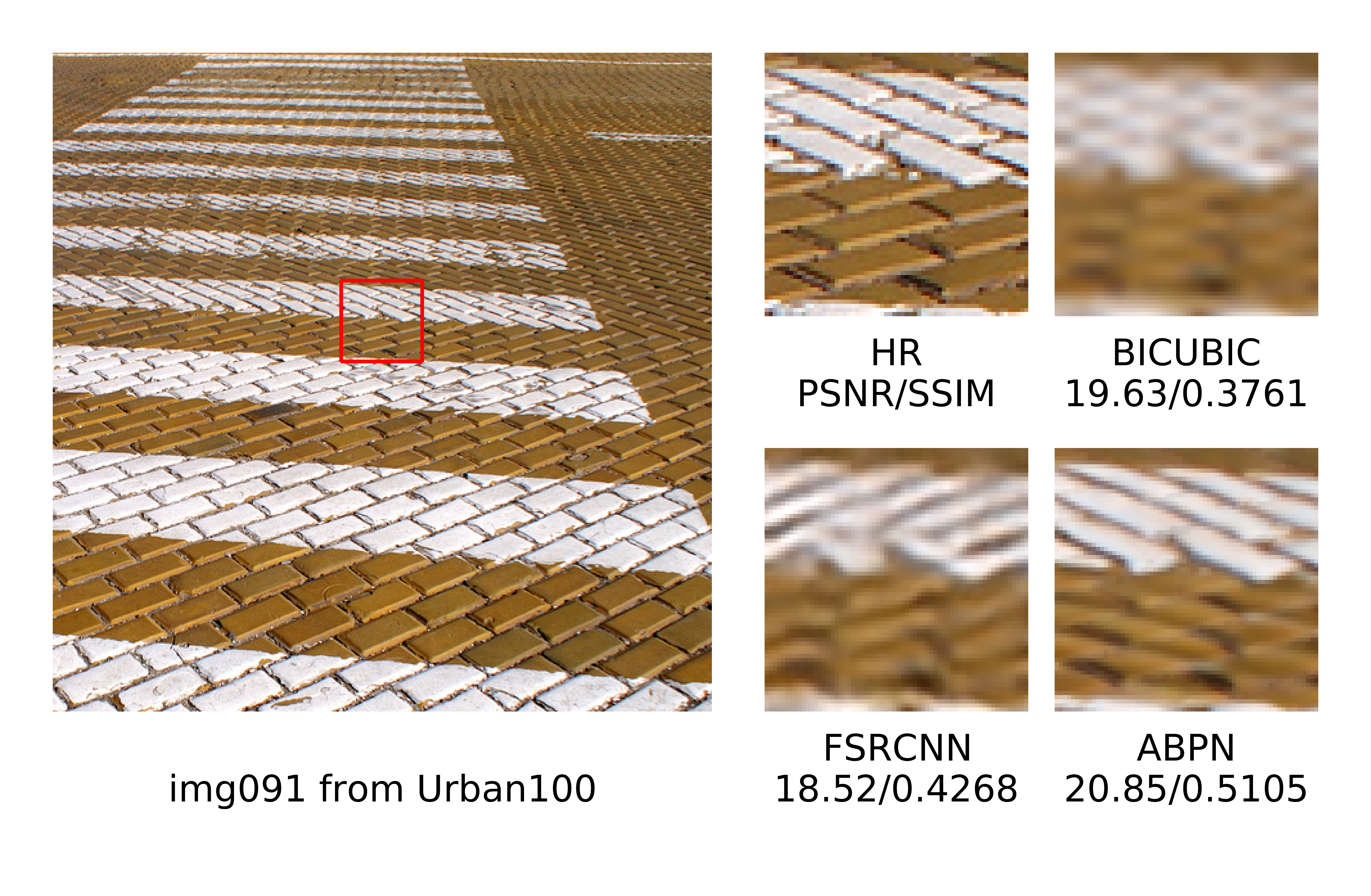} \\
			\end{tabular}
		\end{center}
	    \caption{Visual results on Urban100 of int8 model.}
		\label{fig:visual}	
	\end{figure*}

We show visual comparison of our int8 quantized model and int8 quantized FSRCNN in Fig.~\ref{fig:visual}. Our methods can reconstruct more textures and faithfully edges, which demonstrates the superiority of our proposed ABPN.

\section{Conclusion}
We propose an efficient network called anchor-based plain net (ABPN) for INT8 quantization. The key component is anchor-based residual learning (ABRL), which realize the same functionality of image-space residual learning while being as fast as feature-space residual learning. Our INT8 quantization network can achieve nearly the same performance as original floating-point network.
\clearpage
{\small
\bibliographystyle{ieee_fullname}
\bibliography{Last}
}

\end{document}